\documentclass{article}
\usepackage{graphicx} 
\usepackage{authblk}
\usepackage{amsmath}
\usepackage{threeparttable}
\usepackage{pgfplots}
\usepackage{multirow}
\usepackage{array}
\pgfplotsset{compat=1.18}
\usepackage{caption}
\usepackage{subcaption}
\usepackage{adjustbox}
\usepackage{tabularx}
\usepackage{hyperref}
\usepackage{float}
\usepackage{newunicodechar}
\newunicodechar{−}{\ensuremath{-}}
\usepackage{booktabs}
\usepackage{slashed,epsfig,amsmath,amssymb,enumitem} \usepackage[papersize={8.5in,11in}]{geometry}
   \usepackage{bm}
\usepackage{graphicx}
\newcommand{\be}{\begin{equation}}
\newcommand{\ee}{\end{equation}}
\usepackage[utf8]{inputenc}
\title{Dynamical Dark Energy at Late Time $\Lambda$CDM}

\author[1 2]{J. W. Moffat}

\author[1 3]{E. J. Thompson}

\affil[1]{Perimeter Institute for Theoretical Physics, Waterloo, Ontario N2L 2Y5, Canada}

\affil[2]{Department of Physics and Astronomy, University of Waterloo, Waterloo,
Ontario N2L 3G1, Canada}

\affil[3]{Department of Physics and Astronomy, Trent University, Peterborough, 
Ontario K9L 0G2, Canada}

\begin{document}
\maketitle


\begin{abstract}
We investigate the dynamical properties of dark energy through a detailed analysis of its equation of state parameter $w(z)$ as a function of redshift. We derive a general expression for $w(z)$ from the Friedmann-Lemaître-Robertson-Walker (FLRW) equations, establishing a direct relationship between the dark energy equation of state and the observable Hubble parameter $H(z)$ and its derivative. Using the relation $w(z)=−1+[2(1+z)/3H(z)]dH/dz$, we develop an approximation method valid for $z\lesssim 1$ that accounts for the changing balance between matter and dark energy contributions to cosmic expansion. We compare our theoretical framework with recent observational data from the Dark Energy Spectroscopic Instrument (DESI) DR2, analysing how well the commonly used Chevallier-Polarski-Linder (CPL) parametrization $w(z)=-1+w_a(1/(1+z))$ captures the evolution of dark energy. Our results indicate that the dark energy equation of state exhibits a monotonic evolution with redshift, transitioning from deceleration to acceleration around $z\approx 0.7$. Notably, our predicted $w_{DE}$ remains greater than −1 across all redshifts, avoiding phantom energy scenarios that would violate the null energy condition. This work demonstrates how precise measurements of cosmic expansion history can constrain the nature of dark energy and provides a framework for testing dynamical dark energy models against current and future cosmological observations.

\end{abstract}

\section{Introduction}

The discovery of the accelerated expansion of the universe has transformed modern cosmology, indicating the existence of a mysterious component known as dark energy. While the standard cosmological model attributes cosmic acceleration to a cosmological constant $\Lambda$ with a constant equation of state $w = - 1$, the physical origin and precise nature of dark energy remain elusive. This uncertainty has motivated extensive theoretical and observational investigations into models where the dark energy equation of state, $w$, is allowed to evolve with time or redshift. Distinguishing between a true cosmological constant and dynamical dark energy is one of the foremost goals of precision cosmology.

A powerful way to probe the properties of dark energy is through its impact on the expansion rate of the universe, described by the Hubble parameter $H(z)$. The Friedmann-Lemaître-Robertson-Walker (FLRW) metric, which assumes homogeneity and isotropy on large scales, provides the mathematical foundation for describing the evolution of the universe within general relativity. The Friedmann equations relate the energy content of the universe to its expansion rate, allowing the effects of dark energy to be inferred from measurements of $H(z)$.

A particularly useful approach is to recast the dark energy equation of state in terms of the Hubble parameter and its time derivative, leading to the relation  $w = - 1 -2\dot H/3H^2$. This expression provides a direct link between observations of cosmic expansion and the phenomenology of dark energy. To compare theoretical predictions to data, it is common to employ the Chevallier-Polarski-Linder (CPL)~\cite{CPL} parameterization, $w(z) = w_0 +w_a(z/(1+z))$, which captures possible redshift evolution of the equation of state with only two parameters.

The latest generation of large-scale structure surveys, such as the Dark Energy Spectroscopic Instrument (DESI), provides measurements of the expansion history of the universe through baryon acoustic oscillations (BAO) and other probes \cite{DESI2025BAO, DESI2025LyA, DESI2025DE}. These observations enable stringent constraints on the time and redshift dependence of $w(z)$, bridging theoretical developments and data. In this work, we derive the dynamical equation of state for dark energy from the FLRW equations, analyse its form in terms of observable quantities, and examine its compatibility with the CPL parameterization using recent DESI results. Our study demonstrates the utility of this approach and discusses its implications for the nature of dark energy in light of forthcoming observational constraints.

\section{\texorpdfstring{$\Lambda$}{TEXT}CDM model and transition to dynamical dark energy}

The Friedmann-Lemaître-Robertson-Walker (FLRW) metric describes a homogeneous and isotropic universe with the line element:
\be
ds^2 = - dt^2 + a^2(t) \left[ \frac{dr^2}{1 - k r^2} + r^2 (d\theta^2 + \sin^2\theta \, d\phi^2) \right],
\ee
where $t$ is cosmological time, $a(t)$ is the scale factor at time $t$, $k$ is the spatial curvature parameter: $k=0$ for a spatially flat universe, $k = + 1$ closed universe spherical geometry, $k=-1$ open universe hyperbolic geometry. The FLRW metric describes a homogeneous and isotropic universe. The Friedmann equations (setting natural units, $c=1$) are:
\be
\left(\frac{\dot{a}}{a}\right)^2 = H^2 = \frac{8\pi G}{3}\rho - \frac{k}{a^2} + \frac{\Lambda}{3}
\ee
\be
\frac{\ddot{a}}{a} = -\frac{4\pi G}{3}(\rho + 3p) + \frac{\Lambda}{3},
\ee
where $H=\dot a/a$ is the Hubble parameter, $\rho$ is the total energy density, p is the total pressure, $k =0$ for a spatially flat universe, and  $\Lambda$ is the cosmological constant. The continuity equation expresses energy conservation:
\be
\dot{\rho} + 3H(\rho + p) = 0,
\ee
and we assume an equation of state given by
\be
p = w\rho.
\ee
From the continuity equation we obtain:
\be
\dot{\rho} = -3H\rho(1 + w).
\ee
This leads to the equation:
\be
\frac{3\dot{H}}{4\pi G} = -\frac{3H^2}{8\pi G}(1 + w).
\ee
Rearranging and solving for $w$, we get
\be
w = -1 - \frac{2\dot{H}}{3H^2}.
\ee
This is the equation of state for the total density $\rho$ and pressure $p$. $w > -1$ corresponds to $\dot{H} < 0$, while phantom energy with $w < -1$ corresponds to $\dot{H} > 0$.
The Friedmann equation with $k=0$ is given by:
\be
H^2(z) = \frac{8\pi G}{3} [\rho_m(z) + \rho_r(z) + \rho_{DE}(z)],
\ee
where $\rho_m$, $\rho_r$ and $\rho_{DE}$ are the matter density, radiation density and dark energy density, respectively. Given cosmological observations, $\rho_m(z)$ and $\rho_r(z)$  for matter and radiation evolve in known ways:
\be
\rho_m(z) = \rho_{m0} (1+z)^3,
\ee
\be
\rho_r(z) = \rho_{r0} (1+z)^4.
\ee
It follows that
\be
\rho_{DE}(z) = \frac{3H^2(z)}{8\pi G} - \rho_m(z) - \rho_r(z).
\ee
The dark energy pressure can be found from the acceleration equation.

Subtracting the known components pressure $p_m$ and $p_r=\rho_r/3$ from the total gives
\be
p_{DE}(z) = - \left(\frac{2\dot{H}(z) + 3H^2(z)}{8\pi G} + p_m(z) + p_r(z)\right).
\ee
Once you have $\rho_{DE}$ and $p_{DE}$, we have
\be
w_{DE} =\frac{p_{DE}}{\rho_{DE}}.
\ee
This formula will then give you the dark energy equation of state, not the total equation of state.

For late times or $z < 1$, radiation is negligible:
\be
H^2 = \frac{8\pi G}{3}[\rho_m + \rho_{DE}]
\ee
We can now obtain the dark energy density and pressure:
\be
\rho_{DE} = \frac{3H^2}{8\pi G} - \rho_m,
\ee
\be
p_{DE} = -\frac{1}{8\pi G}(2\dot{H} + 3H^2).
\ee
This gives the dark energy equation of state:
\be
w_{DE} = \frac{p_{DE}(z)}{\rho_{DE}(z)} = -2\dot{H} - \frac{3H^2}{3H^2 - 8\pi G \rho_m}
\ee
For the dark energy dominated and matter negligible case $\rho_{DE} > \rho_m$, we have at late time for $z < 1$:
\be
w_{DE} = - 1 -\frac{2\dot H}{3H^2}.
\ee
We recall the redshift-time relation:
\be
1+z = \frac{1}{a(t)}.
\ee
We have
\be
\frac{dz}{dt} = -(1+z)H,
\ee
leading to $w(z)$ as a function of redshift:
\be
\label{w(z)}
w_{DE}(z) = -1 + \frac{2(1+z)}{3H(z)}\frac{dH}{dz}.
\ee
In the case where dark energy dominates, the ratio simplifies and matches the formula for $w(z)$ via $\dot H$ and $H$ and matches $w_{DE}$ in the dark energy dominated era; otherwise, you must subtract the matter/radiation terms first.

We need to calculate:
\be
w_{DE} = -1 +w_H(z),
\ee
where
\be
w_H(z) = \frac{2(1+z)}{3H(z)}\frac{dH(z)}{dz}.
\ee
In general, $H(z)$ depends on the evolution of dark energy $w_{DE}(z)$ itself. To proceed if $w_{DE}(z)$ is not known a priori, but you know $\Omega_m$ and are working in the regime $z\lesssim 1$, where dark energy and matter are both comparable, but matter is not entirely dominant, we can calculate $w_{DE}$ as an approximate result. In the standard $\Lambda$CDM case $w_{DE}=-1$ and for flat $\Lambda$CDM:
\be
H(z) = H_0 \sqrt{\Omega_m (1+z)^3 + \Omega_\Lambda},
\ee
with 
\be
\Omega_\Lambda = 1-\Omega_m.
\ee
Differentiating this gives
\be
\frac{dH(z)}{dz} = \frac{3H_0(1+z)^2}{2[\Omega_m(1+z)3+\Omega_\Lambda]^{1/2}}.
\ee
Recall that 
\be
H(z)^2 = H_0^2[\Omega_m (1+z)^3 + \Omega_\Lambda].
\ee
We obtain the result:
\be
w_H(z) = \frac{\Omega_m(1+z)^3}{\Omega_m (1+z)^3 + 1 - \Omega_m}.
\ee
We can now approximate $w_{DE}$ for $z\lesssim 1$:
\be
\label{wz2}
w_{DE}(z)\approx - 1 + \frac{\Omega_m(1+z)^3}{\Omega_m (1+z)^3 + 1 - \Omega_m}.
\ee
This formula does not represent the dark energy equation of state in a general cosmology. It represents what you would infer for $w_{DE}(z)$, if you apply the formula derived for dark energy to a universe that has a cosmological constant $w = -1$. We will use the formula assuming that $w_{DE}(z)$ varies with redshift. The formula is valid as an approximation for $z\lesssim 1$, because the formula for $w_{DE}(z)$ is accounting for the changing balance between matter and dark energy contributions to the expansion rate, when $\Omega_m(1+z)^3$ is not negligible compared to $\Omega_\Lambda$. In the redshift range $z\lesssim 1$, both matter and dark energy make significant contributions to the cosmic energy density, but neither completely dominates. The approximation method accounts for this changing balance, which is crucial for accurate modelling.

One can use this as an approximation, if deviations from $ w = -1$ are small. For modest $z < 1$, this approximation is valid given the current observational evidence that $w_{DE}$ changes slowly with $z$.

The deceleration parameter $q$ is defined as:
\begin{equation}
    q = -\frac{a\ddot a}{\dot a^2 }.
\end{equation}

From the Friedmann equations in a universe containing matter and dark energy in a spatially flat universe, the density parameters are defined as: 
$\Omega_m=8\pi G\rho_m/3H^2$, $\Omega_\Lambda=\Lambda/3H^2$ and $\Omega_m+\Omega_\Lambda=1$. For a dark energy equation of state $w_\Lambda=-1$, we obtain for the deceleration parameter:
\be
q = \frac{\Omega_m}{2} - \Omega_\Lambda.
\ee
The universe transitions from deceleration to acceleration when $q=0$. When $\Omega_\Lambda > \Omega_m/2$, we have $q < 0$, indicating an accelerating universe. Using the current values $\Omega_m,0\sim 0.3$ and $\Omega_\Lambda,0\sim 0.7$, the transition from deceleration to acceleration occurred at a redshift of approximately $z\sim 0.7$, or about 6-7 billion years ago.

The Chevallier-Polarski-Linder (CPL)~\cite{CPL} parametrization is given by
\be
w(z) = w_0 + w_a\left(\frac{z}{1+z)}\right),
\label{eq:CPL}
\ee
where $w_0$  is the present-day value at $z=0$, and $w_a$ describes its evolution with redshift $z$. To fit to the CPL formula, we compute $w_{DE}(z)$ for a range of $z$ to give a $w(z)$. We fit the resulting $w(z)$ values in the range of interest $0<z<1$ to the CPL formula. The CPL is an approximation; it captures smooth, slowly-varying behaviours.

\section{DESI data comparison}

The Dark Energy Spectroscopic Instrument (DESI) gives us some of the most precise measurements of the cosmic expansion history using baryon acoustic oscillations (BAO) observations. In this section, we will analyse how these measurements compare to our dynamical dark energy framework.

The DESI DR2 data release makes cosmological data available. The raw spectroscopic and redshift data are presented in a parametrized form, employing the two-parameter $w_0$ and $w_a$ parametrization \cite{CPL} ~\cite{Linder}. This parametrization of the data allows for phantom energy with $w(z) < - 1$. There are other parametrization of the data that can give alternative descriptions of the equation of state $w(z)$ that do not permit phantom energy $w(z) < -1$~\cite{Linder, moresco2016, linder2008}.

We use (\ref{w(z)}) taking the $\Omega_m$ priors provided by the DESI measurements and compute for various redshifts how our model deviates from $\Lambda$CDM. We will compare the values obtained from DESI DR2 with the values we obtain, and aim to show that using our approximation method to obtain $w_{DE}(z)$, we can predict the deviation from the standard $\Lambda$CDM equation of state.

For our analysis, we will calculate $w_{DE}(z)$ at various redshifts $(z)$ and compare them to the DESI DR2 data for $w(z)$ and explore how our model compares to the DESI data.

The priors we take into account are given in table 1\cite{DESI2025BAO, DESI2025LyA, DESI2025DE}.

\begin{table}[ht]
\centering
\resizebox{\textwidth}{200pt}{%
\begin{tabular}{lccccc}
\hline
Model/Dataset & $\Omega_m$ & $H_0$ [km s$^{-1}$ Mpc$^{-1}$] & $10^3\Omega_K$ & $w$ or $w_0$ & $w_a$ \\
\hline
\multicolumn{6}{l}{\textbf{$\Lambda$CDM}} \\
CMB & $0.3169\pm0.0065$ & $67.14\pm0.47$ & --- & --- & --- \\
DESI & $0.2975\pm0.0086$ & --- & --- & --- & --- \\
DESI+BBN & $0.2977\pm0.0086$ & $68.51\pm0.58$ & --- & --- & --- \\
DESI+BBN+$\theta_*$ & $0.2967\pm0.0045$ & $68.45\pm0.47$ & --- & --- & --- \\
DESI+CMB & $0.3027\pm0.0036$ & $68.17\pm0.28$ & --- & --- & --- \\
\hline
\multicolumn{6}{l}{\textbf{$\Lambda$CDM+$\Omega_K$}} \\
CMB & $0.354^{+0.020}_{-0.023}$ & $63.3\pm2.1$ & $-10.7^{+6.4}_{-5.3}$ & --- & --- \\
DESI & $0.293\pm0.012$ & --- & $25\pm41$ & --- & --- \\
DESI+CMB & $0.3034\pm0.0037$ & $68.50\pm0.33$ & $2.3\pm1.1$ & --- & --- \\
\hline
\multicolumn{6}{l}{\textbf{$w$CDM}} \\
CMB & $0.203^{+0.017}_{-0.060}$ & $85^{+10}_{-6}$ & --- & $-1.55^{+0.17}_{-0.37}$ & --- \\
DESI & $0.2969\pm0.0089$ & --- & --- & $-0.916\pm0.078$ & --- \\
DESI+Pantheon+ & $0.2976\pm0.0087$ & --- & --- & $-0.914\pm0.040$ & --- \\
DESI+Union3 & $0.2973\pm0.0091$ & --- & --- & $-0.866\pm0.052$ & --- \\
DESI+DESY5 & $0.2977\pm0.0091$ & --- & --- & $-0.872\pm0.039$ & --- \\
DESI+CMB & $0.2927\pm0.0073$ & $69.51\pm0.92$ & --- & $-1.055\pm0.036$ & --- \\
DESI+CMB+Pantheon+ & $0.3047\pm0.0051$ & $67.97\pm0.57$ & --- & $-0.995\pm0.023$ & --- \\
DESI+CMB+Union3 & $0.3044\pm0.0059$ & $68.01\pm0.68$ & --- & $-0.997\pm0.027$ & --- \\
DESI+CMB+DESY5 & $0.3098\pm0.0050$ & $67.34\pm0.54$ & --- & $-0.971\pm0.021$ & --- \\
\hline
\end{tabular}
}
\label{tab:cosmo_part1}
\end{table}

\begin{table}[ht]
\centering
\resizebox{\textwidth}{200pt}{%
\begin{tabular}{lccccc}
\hline
Model/Dataset & $\Omega_m$ & $H_0$ [km s$^{-1}$ Mpc$^{-1}$] & $10^3\Omega_K$ & $w$ or $w_0$ & $w_a$ \\
\hline
\multicolumn{6}{l}{\textbf{$w_0w_a$CDM}} \\
CMB & $0.220^{+0.019}_{-0.078}$ & $83^{+20}_{-6}$ & --- & $-1.23^{+0.44}_{-0.61}$ & $<-0.504$ \\
DESI & $0.352^{+0.041}_{-0.018}$ & --- & --- & $-0.48^{+0.35}_{-0.17}$ & $<-1.34$ \\
DESI+Pantheon+ & $0.298^{+0.025}_{-0.011}$ & --- & --- & $-0.888^{+0.055}_{-0.064}$ & $-0.17\pm0.46$ \\
DESI+Union3 & $0.328^{+0.019}_{-0.014}$ & --- & --- & $-0.70\pm0.11$ & $-0.99\pm0.57$ \\
DESI+DESY5 & $0.319^{+0.017}_{-0.011}$ & --- & --- & $-0.781^{+0.067}_{-0.076}$ & $-0.72\pm0.47$ \\
DESI+($\theta_*$,$\omega_b$,$\omega_{bc}$)CMB & $0.353\pm0.022$ & $63.7^{+1.7}_{-2.2}$ & --- & $-0.43\pm0.22$ & $-1.72\pm0.64$ \\
DESI+CMB(nolensing) & $0.352\pm0.021$ & $63.7^{+1.7}_{-2.1}$ & --- & $-0.43\pm0.21$ & $-1.70\pm0.60$ \\
DESI+CMB & $0.353\pm0.021$ & $63.6^{+1.6}_{-2.1}$ & --- & $-0.42\pm0.21$ & $-1.75\pm0.58$ \\
DESI+CMB+Pantheon+ & $0.3114\pm0.0057$ & $67.51\pm0.59$ & --- & $-0.838\pm0.055$ & $-0.62^{+0.22}_{-0.19}$ \\
DESI+CMB+Union3 & $0.3275\pm0.0086$ & $65.91\pm0.84$ & --- & $-0.667\pm0.088$ & $-1.09^{+0.31}_{-0.27}$ \\
DESI+CMB+DESY5 & $0.3191\pm0.0056$ & $66.74\pm0.56$ & --- & $-0.752\pm0.057$ & $-0.86^{+0.23}_{-0.20}$ \\
DESI+DESY3(3$\times$2pt)+Pantheon+ & $0.3140\pm0.0091$ & --- & --- & $-0.870\pm0.061$ & $-0.46^{+0.33}_{-0.29}$ \\
DESI+DESY3(3$\times$2pt)+Union3 & $0.333\pm0.012$ & --- & --- & $-0.68\pm0.11$ & $-1.09^{+0.48}_{-0.39}$ \\
DESI+DESY3(3$\times$2pt)+DESY5 & $0.3239\pm0.0092$ & --- & --- & $-0.771\pm0.068$ & $-0.82^{+0.38}_{-0.32}$ \\
\hline
\multicolumn{6}{l}{\textbf{$w_0w_a$CDM+$\Omega_K$}} \\
DESI & $0.357^{+0.041}_{-0.030}$ & --- & $-2\pm56$ & $-0.45^{+0.33}_{-0.17}$ & $<-1.43$ \\
DESI+CMB+Pantheon+ & $0.3117\pm0.0056$ & $67.62\pm0.60$ & $1.1\pm1.3$ & $-0.853\pm0.057$ & $-0.54\pm0.22$ \\
DESI+CMB+Union3 & $0.3273\pm0.0086$ & $65.98\pm0.86$ & $0.6\pm1.3$ & $-0.678\pm0.092$ & $-1.03^{+0.33}_{-0.29}$ \\
DESI+CMB+DESY5 & $0.3193\pm0.0056$ & $66.82\pm0.58$ & $0.8\pm1.3$ & $-0.762\pm0.060$ & $-0.81\pm0.24$ \\
\hline
\end{tabular}
}
\caption{Summary table of cosmological parameter constraints from DESI DR2 BAO (labelled in the table as ‘DESI’) in combination with external data sets and priors, in $\Lambda$CDM and various extended models. Results quoted for all parameters are the marginalized posterior means and 68\% credible intervals in each case where two sided constraints are possible, or the 68\% upper limits when only one-sided constraints are possible.}
\label{tab:cosmo_part2}
\end{table}

\clearpage

Using the $\Omega_m$ priors  in a flat universe ($\Omega_K=0$) we obtain $w_{DE(z)}$ in Fig.~\ref{fig:1,-1}.

\begin{figure}[ht]
    \centering    \includegraphics[width=1\textwidth]{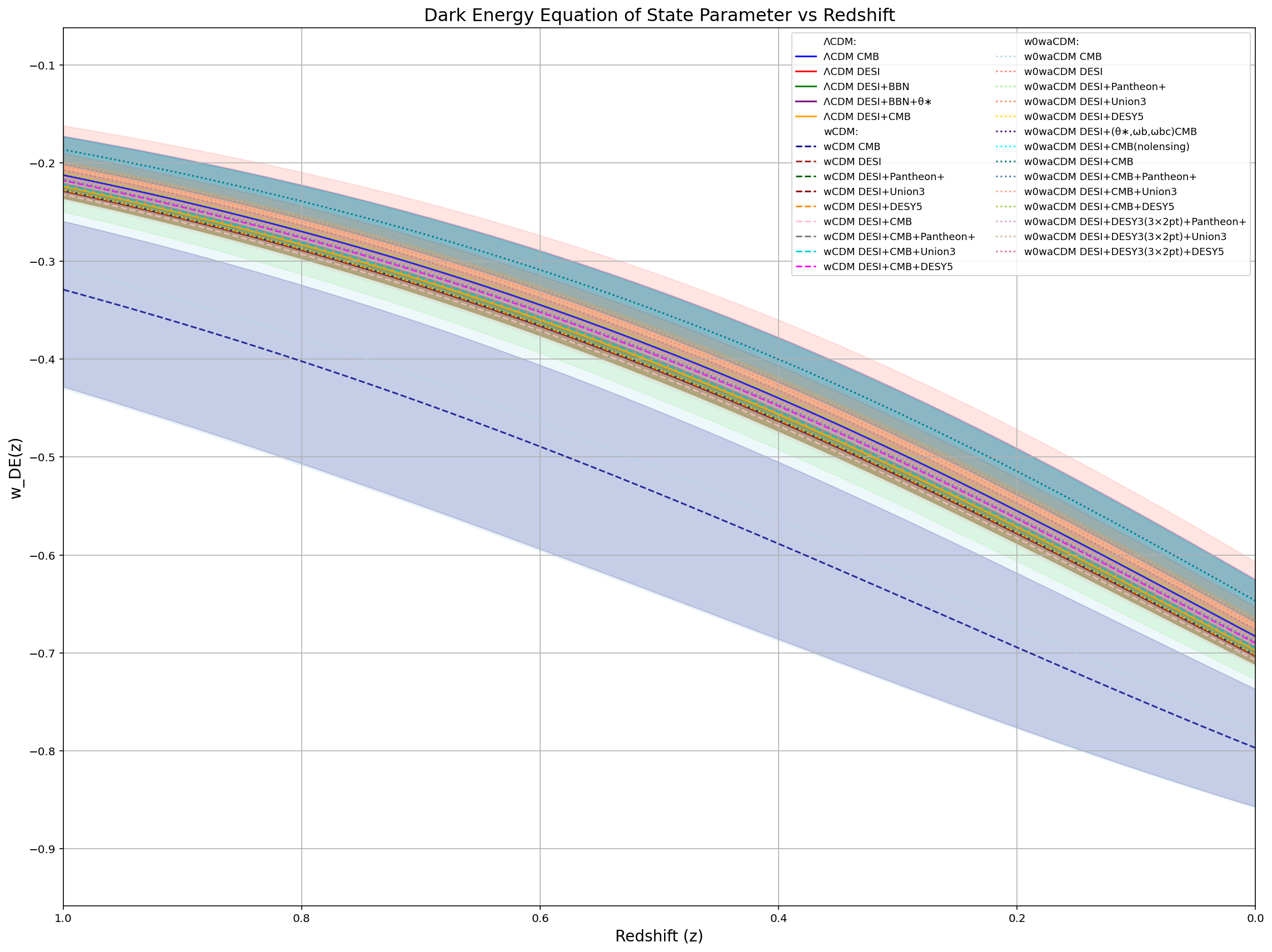}
    \caption{Dark energy equation of state $w_{DE}(z)$ plotted for various $\Omega_m$ priors.}
    \label{fig:1,-1}
\end{figure}

\clearpage

From the three models ($\Lambda CDM,,wCDM, w_0w_aCDM$) we take the two $\Omega_m$ priors with the tightest errors and obtain Fig.~\ref{fig:1,-1-b}.

\begin{figure}[ht]
    \centering    \includegraphics[width=1\textwidth]{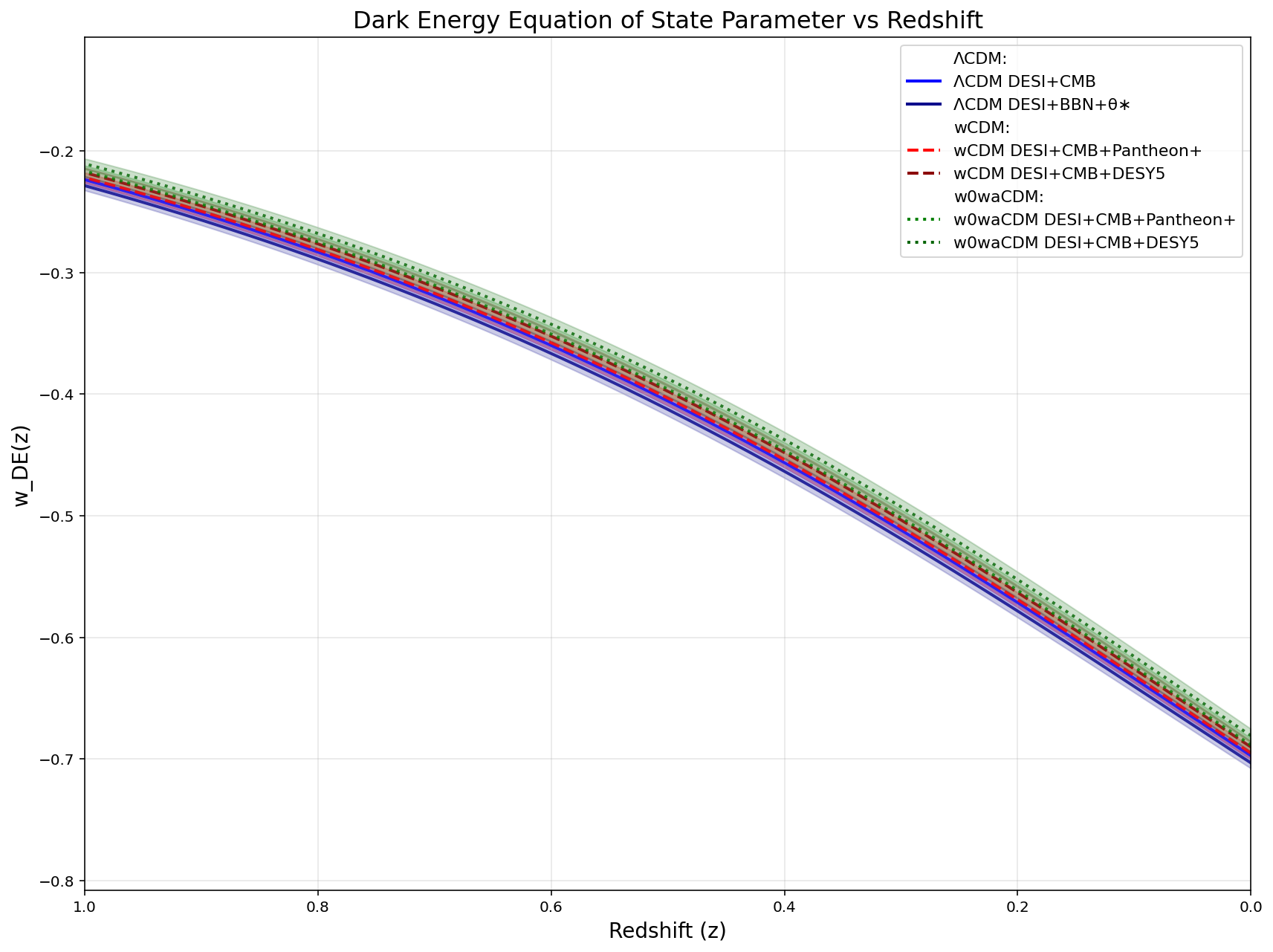}
    \caption{Dark energy equation of state $w_{DE}(z)$ plotted for the $\Omega_m$ priors with the tightest errors.}
    \label{fig:1,-1-b}
\end{figure}

\clearpage

We see at redshift $z = 0$, the majority of best-fit curves across all models and datasets converge around $w_{\text{DE}}(0) \approx -0.7$, with remarkably small uncertainties (typically on the order of  $\pm 0.05$). This constitutes a statistically significant deviation from the $\Lambda$CDM prediction of $w = -1$, which assumes a constant cosmological constant throughout cosmic history. Such a consistent departure strongly suggests that a true cosmological constant may not fully account for the observed acceleration of the universe. Instead, the data increasingly favours dynamical dark energy models. The precision of DESI measurements when combined with CMB, Pantheon+, and BBN+$\theta_*$ datasets, sharply constrains the expansion history, making this deviation difficult to attribute to noise or dataset anomalies. The fact that both $w$CDM and $w_0w_a$CDM models independently converge toward the same trend underscores its robustness across model assumptions and data priors. Importantly, since $w > -1$, the equation of state implies that cosmic acceleration is slower than predicted by $\Lambda$CDM, aligning with recent DESI DR2 results that hint at a weakening of late-time acceleration.

To directly compare our results with DESI, we need to map our model to the CPL parametrization given by equation 33. We need to find $w_0$ and $w_a$ for our model, to do this we will first find $w_0$:
\begin{equation}
    w_0=w_{DE}(0).
\end{equation}
This gives us
\begin{equation}
w_0=-1+\Omega_m.
\end{equation}

Calculating this using our $\Omega_m$ priors yields table 3.

\begin{table}[H]
\centering
\renewcommand{\arraystretch}{1}
\begin{tabularx}{\textwidth}{lcc}
\toprule
\textbf{Model} & $\Omega_m$ & $w_0 = -1 + \Omega_m$ \\
\midrule
$\Lambda$CDM: CMB & 0.3169 & $-0.6831 \pm 0.0065$ \\
$\Lambda$CDM: DESI & 0.2975 & $-0.7025 \pm 0.0086$ \\
$\Lambda$CDM: DESI+BBN & 0.2977 & $-0.7023 \pm 0.0086$ \\
$\Lambda$CDM: DESI+BBN+$\theta_*$ & 0.2967 & $-0.7033 \pm 0.0045$ \\
$\Lambda$CDM: DESI+CMB & 0.3027 & $-0.6973 \pm 0.0036$ \\
$w$CDM: CMB & 0.2030 & $-0.7970 \pm 0.0385$ \\
$w$CDM: DESI & 0.2969 & $-0.7031 \pm 0.0089$ \\
$w$CDM: DESI+Pantheon+ & 0.2976 & $-0.7024 \pm 0.0087$ \\
$w$CDM: DESI+Union3 & 0.2973 & $-0.7027 \pm 0.0091$ \\
$w$CDM: DESI+DESY5 & 0.2977 & $-0.7023 \pm 0.0091$ \\
$w$CDM: DESI+CMB & 0.2927 & $-0.7073 \pm 0.0073$ \\
$w$CDM: DESI+CMB+Pantheon+ & 0.3047 & $-0.6953 \pm 0.0051$ \\
$w$CDM: DESI+CMB+Union3 & 0.3044 & $-0.6956 \pm 0.0059$ \\
$w$CDM: DESI+CMB+DESY5 & 0.3098 & $-0.6902 \pm 0.0050$ \\
$w_0w_a$CDM: CMB & 0.2200 & $-0.7800 \pm 0.0485$ \\
$w_0w_a$CDM: DESI & 0.3520 & $-0.6480 \pm 0.0295$ \\
$w_0w_a$CDM: DESI+Pantheon+ & 0.2980 & $-0.7020 \pm 0.0180$ \\
$w_0w_a$CDM: DESI+Union3 & 0.3280 & $-0.6720 \pm 0.0165$ \\
$w_0w_a$CDM: DESI+DESY5 & 0.3190 & $-0.6810 \pm 0.0140$ \\
$w_0w_a$CDM: DESI+($\theta_*$, $\omega_b$, $\omega_{bc}$)CMB & 0.3530 & $-0.6470 \pm 0.0220$ \\
$w_0w_a$CDM: DESI+CMB (no lensing) & 0.3520 & $-0.6480 \pm 0.0210$ \\
$w_0w_a$CDM: DESI+CMB & 0.3530 & $-0.6470 \pm 0.0210$ \\
$w_0w_a$CDM: DESI+CMB+Pantheon+ & 0.3114 & $-0.6886 \pm 0.0057$ \\
$w_0w_a$CDM: DESI+CMB+Union3 & 0.3275 & $-0.6725 \pm 0.0086$ \\
$w_0w_a$CDM: DESI+CMB+DESY5 & 0.3191 & $-0.6809 \pm 0.0056$ \\
$w_0w_a$CDM: DESI+DESY3 (3$\times$2pt)+Pantheon+ & 0.3140 & $-0.6860 \pm 0.0091$ \\
$w_0w_a$CDM: DESI+DESY3 (3$\times$2pt)+Union3 & 0.3330 & $-0.6670 \pm 0.0120$ \\
$w_0w_a$CDM: DESI+DESY3 (3$\times$2pt)+DESY5 & 0.3239 & $-0.6761 \pm 0.0092$ \\
\bottomrule
\end{tabularx}
\end{table}

The relationship $w_0 = -1 + \Omega_m$ appears to emerge as a natural parameter degeneracy in specific cosmological parameter combinations, particularly evident in datasets like DESI+CMB+Union3. This correlation likely arises from the way these parameters jointly affect cosmological observables, notably the angular diameter distance and expansion history. The varying degree of agreement across different dataset combinations suggests this relationship becomes more pronounced when certain observational constraints are combined, possibly indicating the presence of underlying physical connections in dark energy and matter interactions that are revealed only when specific geometric and growth of structure probes are analysed together. The better agreement seen in some $w_0w_a$CDM models compared to $w$CDM cases could indicate that time varying dark energy scenarios naturally accommodate this degeneracy, while showing greater tension in CMB-only analyses may reflect the different sensitivity of high redshift observations to the dark energy equation of state. This parameter relationship merits further investigation as a potential insight into dark sector physics or as a useful approximation for cosmological analyses when certain observational probes are combined.

To calculate our model $w_a$, we will use the relationship to the CPL:
\begin{equation}
\frac{dw_{DE}(z)}{dZ}= w_a,
\end{equation}
where $Z = z/1+z$. This gives us
\begin{equation}
\frac{dw_{DE}}{dZ}=\frac{3\Omega_m(1+z)^4(1-\Omega_m)}{[\Omega_m(1+z)^3+1-\Omega_m]}.
\end{equation}
Evaluating this expression at $z=0$ yields
\begin{equation}
\left. \frac{dw_{DE}}{dZ} \right|_{z=0}=3\Omega_m(1-\Omega_m) = w_a.
\end{equation}

We now reconstruct our CPL parametrized $w_{DE}(z)$ model in order to compare with DESI results. We obtain:
\begin{equation}
w_{DE}^{CPL}(z)=-1+\Omega_m+\left(3\Omega_m(1-\Omega_m)\left(\frac{z}{1+z}\right)\right).
\end{equation}
When $\Omega_m \lesssim\Omega_\Lambda$, we obtain a cosmological constant $\Lambda$ and $w_{DE}\sim -1$.

\clearpage

\begin{figure}[H]
    \centering
    \includegraphics[width=0.9\textwidth]{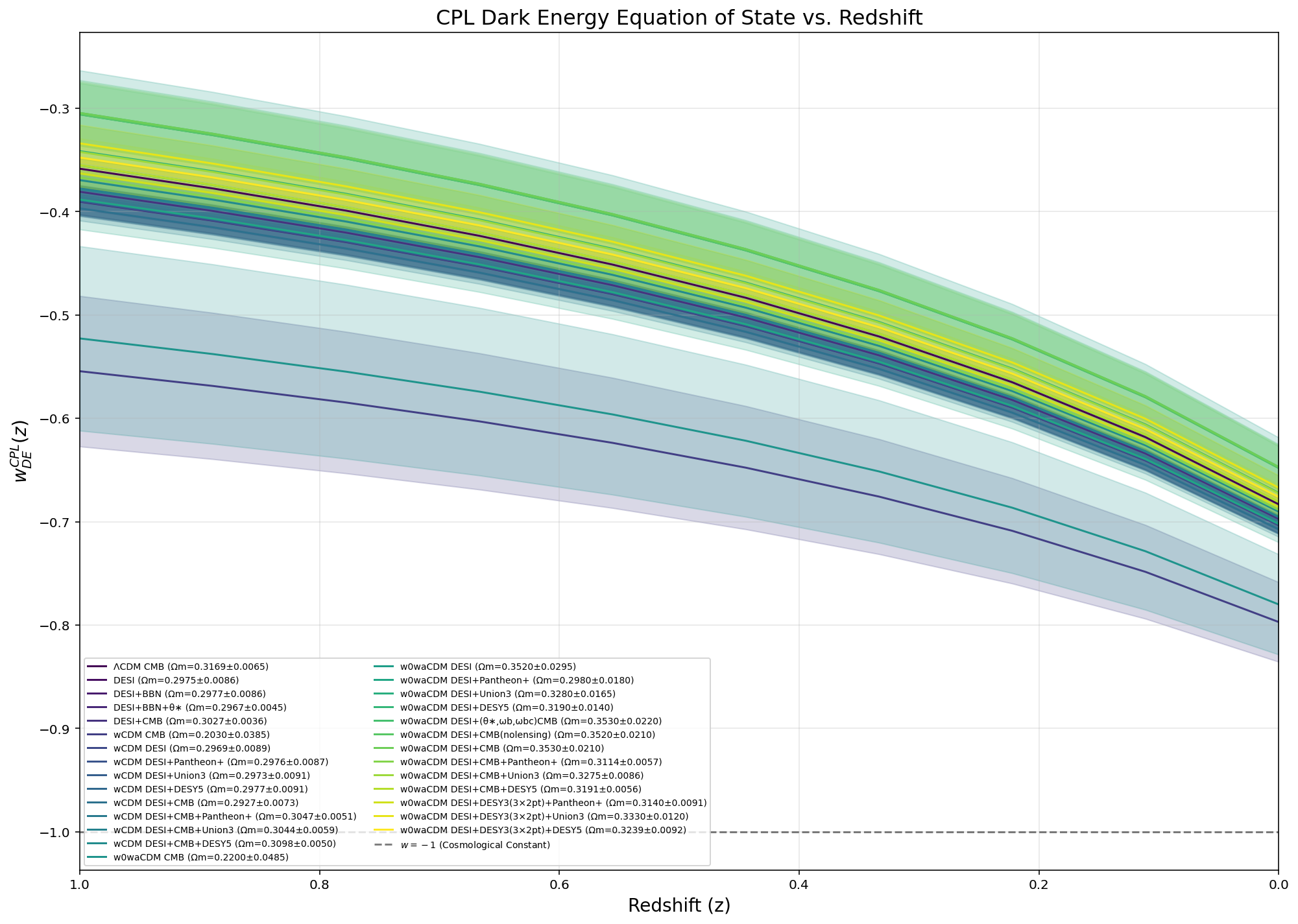}
    \caption{Evolution of the theoretical CPL parametrized dark energy equation of state, $w^{\text{CPL}}_{DE}(z)$.}
    \label{fig:CPLwDE}
\end{figure}

Solving for $0<z<1$, we obtain Fig.~\ref{fig:CPLwDE}.

\begin{figure}[H]
    \centering
    \includegraphics[width=0.8\textwidth]{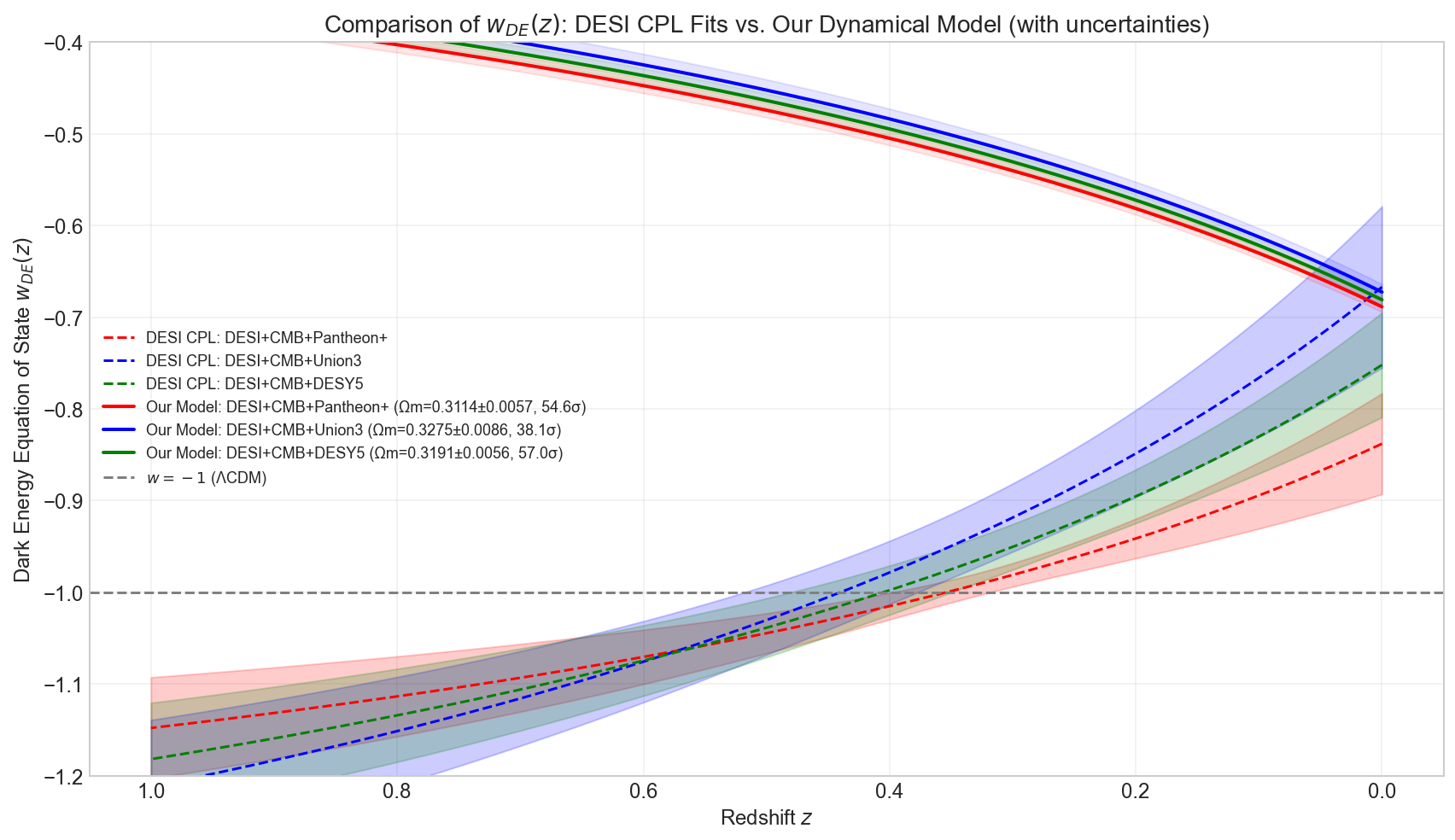}
    \caption{Comparison between the theoretical CPL dark energy equation of state model and DESI DR2 best-fit CPL parametrizations.}
    \label{fig:compare_CPL}
\end{figure}

We will now compare this to the DESI CPL parametrization and obtain Fig.~\ref{fig:compare_CPL}.

We compare our theoretical CPL parametrization of the dark energy equation of state (39) to the CPL fits from DESI DR2 observational data. As shown in Fig. 4, our model yields a smooth, monotonic evolution of $w_{DE}(z)$ with redshift and remains strictly above the cosmological constant boundary $w = -1$ throughout the redshift range $0 \leq z \leq 1$, consistent with a non-phantom dark energy component.

In contrast, the DESI CPL fits suggest a transition into the phantom regime $w < -1$ at intermediate redshifts, with central values deviating from our theoretical curves by up to $ \Delta w \sim 0.1 - 0.2 \quad \text{at} \quad z \sim 0.5. $

The DESI reconstructions imply a more rapid evolution of dark energy than our model predicts. This deviation reflects the sensitivity of the CPL fit to combined datasets with differing systematics and degeneracies in $w_0$ and $w_a$. Our model, being fully determined by $\Omega_m$ and respecting the null energy condition, offers a conservative and physically consistent alternative to parametrized fits that permit phantom crossing. The comparison emphasizes that although observational reconstructions hint at dynamical behaviour, the underlying expansion history may still be well described by non-phantom, $\Omega_m$-driven dynamics.

The Chevallier–Polarski–Linder (CPL) parameterization,
\begin{equation}
w(z) = w_0 + w_a \frac{z}{1 + z},
\tag{\ref{eq:CPL}}
\end{equation}
is widely employed in observational cosmology due to its flexibility and computational convenience. It enables a first-order expansion of the dark energy equation of state around $z = 0$ and remains finite as $z \rightarrow \infty$, which makes it suitable for fitting data across a broad redshift range. Furthermore, it allows analytic integration of the Hubble parameter $H(z)$, simplifying the analysis of large datasets such as those from DESI DR2. However, the CPL form is not derived from any underlying physical theory. Because of this, the CPL form permits behaviours such as phantom crossing ($w < -1$) that violate theoretical consistency conditions, including the null energy condition and stability criteria. In contrast, our model arises directly from the FLRW framework with minimal assumptions, requiring only the matter density parameter $\Omega_m$. As shown in Fig. 5, our prediction for $w_{DE}(z)$ exhibits a smooth, non-phantom evolution, remaining strictly above $w = -1$ for all $z$. The observed deviation between our model and the CPL-based DESI fits may be attributed to the fact that CPL is a purely phenomenological fitting tool, not constrained by physical priors. This suggests that while CPL offers a flexible framework for data interpolation, it may over fit or misrepresent the true physical evolution of dark energy if not properly interpreted.

\section{Future dark energy predictions in \texorpdfstring{$\Lambda$}{TEXT}CDM model}

The weak energy condition (WEC) is given by:
\begin{equation}
    T_{\mu\nu}t^\mu t^\nu = p+\rho\geq0,
\end{equation}

where $\rho\geq0$, and $t^\mu$ is a time like vector.

The strong energy condition (SEC) is given by:

\begin{equation}
    (T_{\mu\nu}-\frac{1}{2}T_{\mu\nu}g_{\mu\nu})t^\mu t^\nu = \rho+3p\geq0,
\end{equation}

where $\rho+p\geq0$.

The null energy condition (NEC) states: 

\begin{equation}
    T_{\mu\nu}k^\mu k^\nu = \rho + p \geq 0.
\end{equation}

These conditions play an important role in GR.
$T_{\mu\nu}$ is the energy-momentum tensor for a perfect fluid with energy density $\rho$ and pressure $p$ and $k^\mu$ is a null vector $k^2=0$. When $w_{DE} < -1$ the dark energy crosses into the phantom dark energy and violates the NEC. This form of dark energy will cause the universe to expand at an ever-increasing rate, leading to a "Big Rip" scenario, where the expansion becomes so violent that it tears apart galaxies and atoms. It can cause instabilities in the vacuum and problems with causality. When we consider dark energy as the vacuum energy with equation of state $w_{DE} = - 1$, we have
\be 
T_{\mu\nu}k^\mu k^\nu = 0.
\ee
This represents the borderline case when the null energy is satisfied. The dark energy with 
$w_{DE} = -1$ is at the boundary of what is allowed by the NEC. This boundary is called the phantom divide in cosmology.

Our prediction for $w_{DE}(z)$ is a monotonically accelerating dark energy expansion, commencing at $z= 0.6 - 0.7$ and continuing accelerated expansion until today's redshift z = 0. This behaviour of $w_{DE}(z)$ is displayed in Fig. 1 and 2 . The predicted $w_{DE}(z)$ never crosses into the phantom energy regime, $w_{DE}(z) < -1$, avoiding a violation of the NEC and preserving the fundamental physical assumptions of GR. 

In Fig. 5 and 6, the evolution of the predicted $w_{DE}(z)$ versus $z$ is displayed. Significantly, $w_{DE}(z)$ asymptotically approaches $w_{DE}(z) = - 1$ avoiding the "Big Rip" scenario.

Now, using equation 30 to calculate $w_{DE}(z)$, we extend our model into the future to explore the predictions it could lead to, using the 28 $\Omega_m$ priors we obtain Fig.~\ref{fig:1,-1-a}.

\begin{figure}[H]
    \centering    \includegraphics[width=1\textwidth]{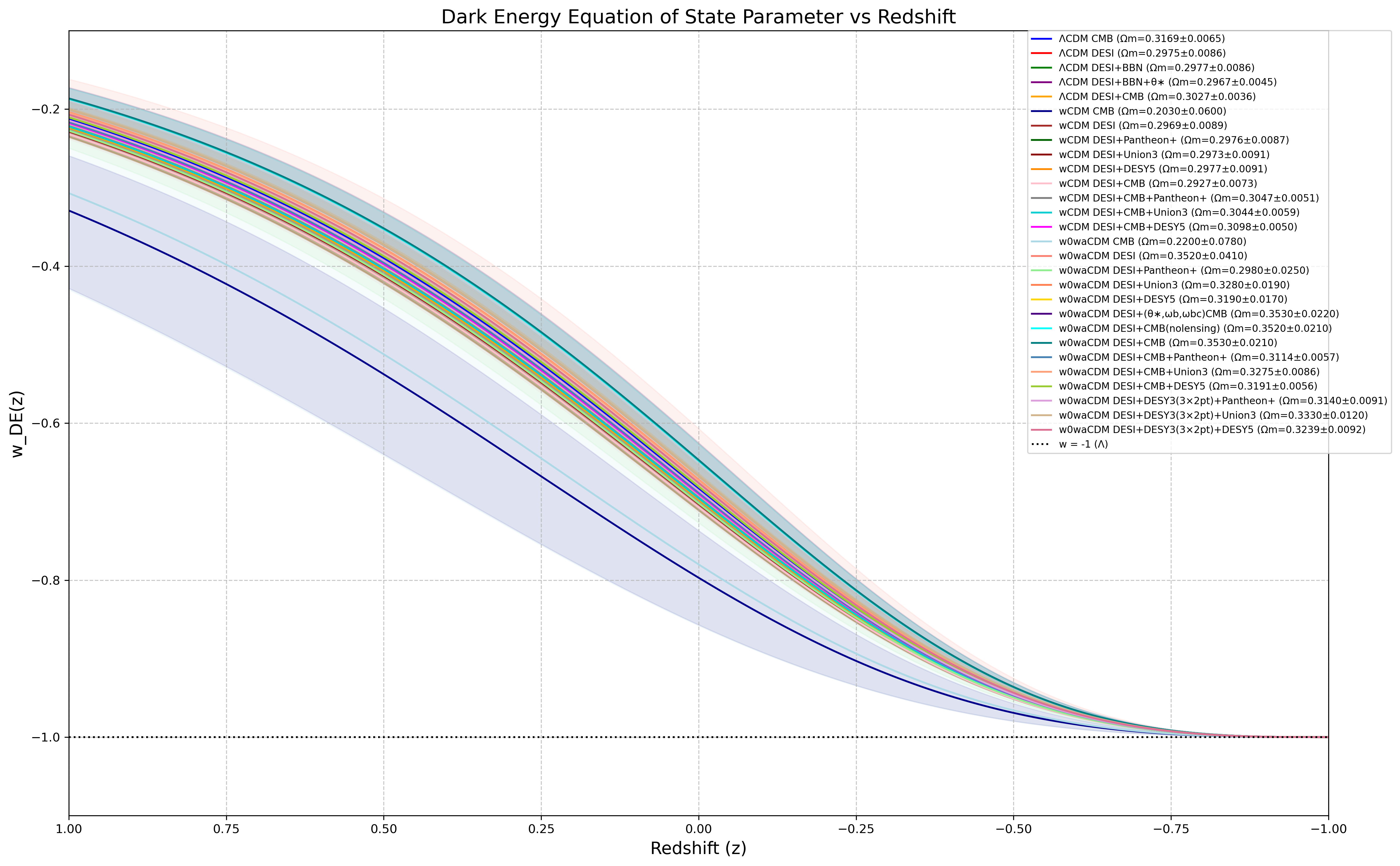}
    \caption{Evolving dark energy equation of state $w_{DE}(z)$ plotted for various $\Omega_m$ priors.}
    \label{fig:1,-1-a}
\end{figure}

\clearpage

For ease of reading, we will display the best two datasets from the three distinct models in Fig.~\ref{fig:1,-1-c}.

\begin{figure}[H]
    \centering    \includegraphics[width=0.8\textwidth]{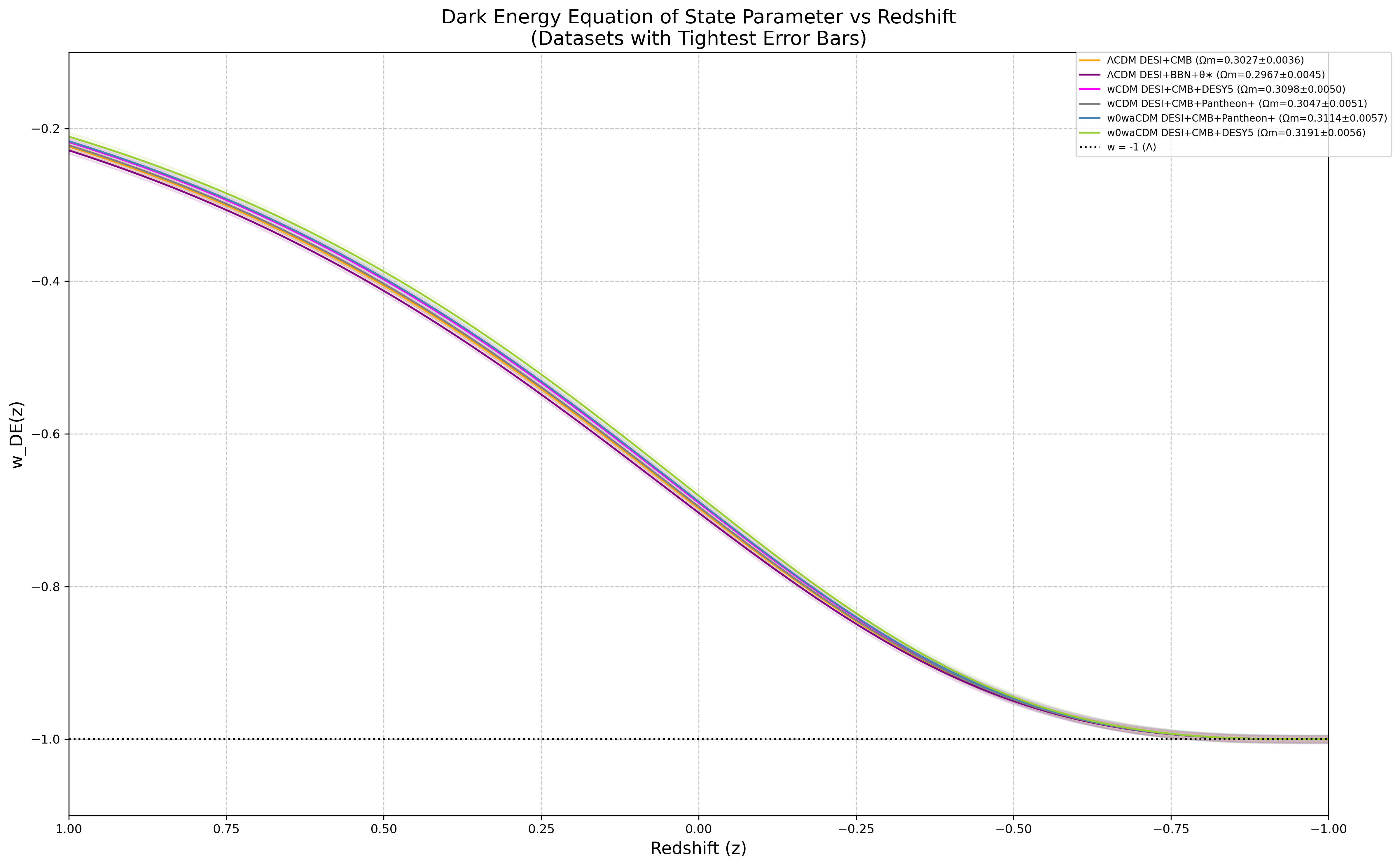}
    \caption{Evolving dark energy equation of state $w_{DE}(z)$ plotted for various $\Omega_m$ priors with the tightest constraints.}
    \label{fig:1,-1-c}
\end{figure}

From our model, we can see that we do not violate the NEC as even at $z \lesssim 0 $, we are always greater than $-1$ and we recover the standard $\Lambda$CDM predictions.
While several DESI-inferred models, particularly those using the $w_0w_a$CDM parametrization, allow for a crossing into the phantom regime at late time redshift ($z > 1$), our reconstructed equation of state $w_{\text{DE}}(z)$ remains strictly greater than $-1$ across the full redshift range considered. This behaviour avoids the problems associated with phantom dark energy, such as violations of the null energy condition and future singularities. As such, our results preserve a $\Lambda$CDM-like asymptotic behaviour at late and future times, despite deviating from $w = -1$ in the present epoch. In this sense, our model both challenges and partially preserves $\Lambda$CDM, it favours dynamical dark energy today, while maintaining theoretical consistency and stability into the future.

\section{Conclusions}

We have explored the relationship between the expansion history of the universe and the dynamical properties of dark energy within the FLRW cosmological framework. By deriving the effective equation of state parameter as $w(z) = -1-2\dot H/3H^2$, we established a direct method to translate measurements of the Hubble parameter and its evolution into constraints on dark energy dynamics. This approach eliminates the need to assume a particular physical model for dark energy, instead enabling a data-driven inference from cosmological observations.
We have explored the dynamics of dark energy through its imprint on the cosmic expansion history, focusing on the relationship between the equation of state parameter $w(z)$ and the Hubble parameter $H(z)$ and its derivative within the FLRW cosmological framework. We derived an explicit, observationally accessible expression for $w(z)$ in terms of $H(z)$, and illustrated how this relation provides an approximate method to probe the nature of dark energy independently of assumptions about its microphysical origin.

The analysis of recent DESI DR2 data within this framework reveals several important findings. First, our derived $w_{DE}$ shows a clear evolution with redshift, with values that remain consistently above $−1$, thus avoiding phantom energy scenarios that would violate the null energy condition. This behaviour is consistent with a universe that transitioned from deceleration to acceleration around $z\approx 0.7$, in agreement with the standard cosmological timeline.

When comparing our results with the CPL parametrization commonly used in cosmological analyses, we find that while this two-parameter model provides a reasonable approximation to the dark energy equation of state for $z<1$, there are subtle deviations that may become increasingly important as observational precision improves. The best-fit values from DESI+CMB+DESY5 data suggest a dark energy component that evolves more rapidly than predicted by the cosmological constant model.

Our extension of the model to negative redshifts and future epochs, indicates that the universe will continue its accelerated expansion without approaching a "Big Rip" scenario, as $w_{DE}$ asymptotically approaches −1 rather than crossing into the phantom regime. This prediction has significant implications for the ultimate fate of our universe.

The methodology presented here can be readily applied to forthcoming data from next-generation surveys, which will provide even more precise measurements of $H(z)$ across a wider range of redshift. Such improvements will enable more stringent tests of dynamical dark energy models and reveal subtle features in $w(z)$ that could point toward the fundamental nature of dark energy.

Our work demonstrates that the combination of theoretical frameworks derived from fundamental physics with high-precision observational data offers an approach to addressing one of the most profound questions in modern cosmology: the nature and evolution of dark energy. While current data is consistent with a dynamically evolving dark energy component, future observations will be crucial in determining whether this evolution represents a genuine departure from the cosmological constant paradigm within the standard $\Lambda$CDM framework.

\section*{Acknowledgments}

We thank Viktor Toth for helpful discussions. We would also like to thank Will Percival, Dustin Lang, and Alex Krolewski for insightful discussions on dark energy and DESI.
Research at the Perimeter Institute for Theoretical Physics is supported by the Government of Canada through Industry Canada and by the Province of Ontario through the Ministry of Research and Innovation (MRI).

\end{document}